\documentstyle[12pt]{article}
\newcommand{\cts}{\int\limits_s^t}
\newcommand{\be}{\begin{equation}}
\newcommand{\ee}{\end{equation}}
\newcommand{\sit}{\sigma^2(t)}
\newcommand{\sik}{\sigma^4(t)}
\newcommand{\elt}{e^{-\lambda t}}
\newcommand{\els}{e^{-\lambda (t\,-\,s)}}
\newcommand{\ela}{e^{-\lambda\tau}}
\newcommand{\llt}{\frac{1\:-\:e^{-\lambda (t\,-\,s)}}{\lambda}}
\newcommand{\yks}{u\:-\:v\elt}
\newcommand{\zes}{z^2\:+\:\sik}
\newcommand{\lit}{\lim\limits_{t\to 0}\frac{1}{t}}
\newcommand{\caa}{\int\limits_{-\infty}^{\infty}}
\newcommand{\fep}{\hat{f}(p)}
\newcommand{\sil}{\frac{\sigma^2}{\lambda}}
\newcommand{\fta}{\cts f(\tau)d\tau}
\newcommand{\sla}{\frac{\sigma}{\lambda}}
\newcommand{\lts}{(\els\:-\:1\:+\:\lambda(t\:-\:s))}
\newcommand{\uou}{{\bf u}_{OU}}
\newcommand{\uout}{\uou(t)}
\newcommand{\sini}{\sigma^2(\infty)}
\newcommand{\sici}{\sigma^4(\infty)}
\newcommand{\lini}{\lim\limits_{t\to\infty}}
\newcommand{\sip}{\frac{\sini}{\pi}}

\begin{document}
\baselineskip18pt
\oddsidemargin0cm
\textwidth15.5cm
\textheight21cm

\title{Ornstein--Uhlenbeck--Cauchy Process}  

\author{Piotr Garbaczewski\\
Institute of  Physics, Pedagogical University \\
PL-65 069 Zielona G\'{o}ra,  Poland \\
and\\
Robert Olkiewicz\\ Institute of Theoretical Physics, University of 
Wroc{\l}aw\\
 PL 50-204 Wroc{\l}aw, Poland}

\maketitle

\noindent
\begin{abstract}
We combine earlier investigations of linear systems with
L\'{e}vy fluctuations [Physica {\bf 113A}, 203, (1982)] with
 recent discussions of L\'{e}vy flights in external force
fields [Phys.Rev. {\bf E 59},2736, (1999)]. We
 give a complete construction of the
 Ornstein-Uhlenbeck-Cauchy process as a
 fully computable model of an anomalous transport and
 a paradigm example of Doob's stable noise-supported
Ornstein-Uhlenbeck process. Despite the nonexistence of all
moments, we determine
 local characteristics (forward drift) of the process,
generators of forward and backward dynamics, relevant
(pseudodifferential) evolution equations. Finally we prove that
this random dynamics is not only mixing (hence ergodic) but also
exact.   The induced nonstationary spatial process is proved to be
Markovian and quite apart from its inherent discontinuity
defines an associated  velocity process  in a probabilistic sense.
\end{abstract}

\section{Motivation}

The casual  understanding of the central limit theorem (in
reference to the  Boltzmann-Gibbs thermostatics), combined with
the need to have clearly specified  the mean  features (moments
and local conservation laws) of  randomly implemented transport,
at or off thermal equlibrium,   resulted in an
obvious predominance of Gaussian laws of probability  and diffusion
processes in typical statistical
(eventually probabilistic, cf. the omnipresence of the Brownian motion
conceptual background)  analysis  of physical phenomena.

Presently, we  observe a   continually growing recognition
of the  profound r\^{o}le (ubiquity, \cite{tsallis}) played by
 non-Gaussian L\'{e}vy distributions (probability laws) in both
 a consistent probabilistic interpretation of various  experimental
 data and in a stochastic modelling of  physical phenomena,
 followed by numerical and realistic experimentation attempts to
 verify (or rather falsify)  probabilistic hypotheses.

 Generically, L\'{e}vy's
 probability laws appear in the context of anomalous diffusions
 (mostly subdiffusions that are  modelled in terms of continuous
 random walks, \cite{shlesinger}).
 On the other hand, under the name of L\'{e}vy flights,
 \cite{shlesinger,metzler} we encounter   stochastic
jump-type  processes which  are explicitly associated with
those distributions.
That allows in
 turn to model quite a variety of  transport  processes,
 cf. \cite{metzler} which are either regarded as  (non)typical
 phenomena of   nonequilibrium statistical physics
 or as manifestations  of a  complex nonlinear  dynamics
 with signatures of chaos,
 yielding  an enhanced diffusion in particular.

We focus our attention on L\'{e}vy flights which  are
considered as possible models of   primordial  noise,
\cite{klaud,olk}.
 (Wiener noise or process  is normally interpreted
to represent a statistical "state of rest" of the random
medium).
Generically, the variance and higher cumulants of those processes
are infinite (nonexistent). There is also  physically more
singular subclass of such processes for which even the
 first moment (mean value) is nonexistent.
Thus we need to relax the limitations of  the standard Gaussian
paradigm:  we  face here  a fundamental problem of
establishing   other means (than variances and mean values)
to characterise statistical properties
of L\'{e}vy processes (fractional moments of Ref. \cite{west} are
insufficient tools in this respect).

Specifically, if a habitual
statistical analysis is performed on any experimentally
available set of frequency data,
there is no obvious method to extract a  reliable  information
about tendencies (local mean values) of
the random dynamics.
Nonexistence of mean values and higher moments may also be interpreted
as the nonexistence of observable (e.g. mean, like drifts or local
currents) regularities of
the dynamics. Moreover, the jump-type processes usually admit
arbitrarily small   jumps (with no lower bound) and finite,
but arbitrarily   large jump sizes (with no upper bound).
Any computer simulation   utilizes both the lower (coarse-graining)
and upper  bound on the jump size, \cite{stanley,mantegna,weron},
and any experimental data collection involves such limitations as well.
Mathematically, that puts us in the framework of standard jump processes
for which the central limit theorem is known to hold true in its
Gaussian version (even if we account for the abnormally slow convergence
to a Gaussian, in view of long tails of the probability distribution,
\cite{stanley}).
Therefore, there is no clear-cut procedures allowing to attribute
an unambigous  statistical interpretation in terms of
 L\'{e}vy processes to  given  phenomenological data. In a drastic
 contrast to a traditional   Gaussian modelling. Mere scaling
 arguments, reflecting the self-similiar patterns of sample  paths,
  are  insufficient as well.

Although no realistic formulation of a
fluctuation-dissipation theorem is possible in that case (nonexistence
of variances), we can
give a meaning to a  theory of L\'{e}vy flights in external force
fields, \cite{metzler}, under a simplifying assumption that
force fields define linear processes with L\'{e}vy
fluctuations. The corresponding velocity processes  were introduced
in  Ref.  \cite{west} ( see also \cite{cac}),  but we shall give
a  complete
construction of the related jump-type stochastic process, together
with a detailed characterization of
the dynamics of induced spatial displacements.
Our strategy is thus substantially different from that
typically followed in the current literature, \cite{metzler}.
For example, the configuration space Langevin equation,
\be \frac{d{\bf x}(t)}{dt}\;=\; {{\bf F}({\bf x})\over {m\gamma }}
\;+\;{\bf \eta }(t)\ee
where $m$ is the mass of transported particle, $\gamma $ stands for
the friction  constant and $\bf \eta $ represents any  conceivable
generalization of the white-noise that employs L\'{e}vy stable
statistics, \cite{metzler}, corroborates a 
tacit assumption that some  kind
 of the the standard    Smoluchowski projection  
(the large friction limit, normally  employed 
 in the Brownian motion context, \cite{chandra})  from
 the phase-space to 
spatial only dynamics   is possible. This is certainly not 
realizable in the non-Gaussian L\'{e}vy  case.

Another delicate question is to settle possible physical origins of
the   spatial noise. That  issue seems to be
conceptually easier to handle  on the velocity/momentum space level.
However, another delicate problem is immediate: in case of the 
Brownian motion (Ornstein-Uhlenbeck process) spatial trajectories were 
by construction differentiable to give meaning to the velocity
concept (even though accelerations were nonexistent anyway,
somewhat conflicting  with the naive but widely spread usage of the
white noise-supported Langevin equation as the second Newton law
analogue).
The conseuqent exploitation  of L\'{e}vy processes with
their intrinsic
discontinuities, seems to set an unresolvable obstacle in this 
(velocity notion) respect.

We shall demonstrate that this 
is not literally the case. For example,  we  can
prove that the spatial random variable  ${\bf x}(t)$ of the
Ornstein-Uhlenbeck-Cauchy 
process cannot possess derivatives in the sense of standard 
mathematical analysis, nonetheless this process has derivatives 
in a weaker, probabilistic sense. Hence, it is legitimate
to interpret  ${\bf u}(t)$ as a velocity  analogue attributed to
the instantaneous (random) location   ${\bf x}(t)$,
though this notion is
more distant from  classical intuitions than the
 velocity variable of the standard Ornstein-Uhlenbeck process
 (not differentiable, hence not yielding  any analogue of a
 Newtonian acceleration).

Our analysis  departs from a  generalization of
 the Ornstein-Uhlenbeck process due to Doob, \cite{doob}, where
a symmetric  L\'{e}vy (stable)  noise was assumed to take place
of the standard Wiener noise.
A complete  description  of a concrete, computable  in full detail
Ornstein-Uhlenbeck-Cauchy process  (with a  familiar
lorentzian as  a probability law for velocity displacements) is our
principal  goal in the
present paper.
In addition we shall pay attention to intrinsic
complications of the random dynamics by investigating a standard
chain of its possible features (ordered with respect to
the complication level): ergodicity, mixing and exactness.

\section{Langevin equation with a linar (harmonic)  force
and Cauchy noise}

The starting point for Ornstein and Uhlenbeck,\cite{ou1,ou2},
 was the dissipative Langevin equation
\be \frac{d{\bf u}}{dt}\;=\;-\lambda{\bf u}(t)\;+\;A(t)\ee
where ${\bf u}(t)$ is a random variable describing the velocity
of a particle, $\lambda>0$ is a
dissipation constant, and $A(t)$ is another random
variable whose probabilistic features  are determined
by the probability distribution of ${\bf u}(t)$, which
is assumed to satisfy a concrete law when $t\to\infty$.
Because ${\bf u}(t)$ may have no time derivative, equation (2)
was soon replaced by another one, namely
\be d{\bf u}(t)\;=\;-\lambda{\bf u}(t)dt\;+\;
dB(t),\quad {\bf u}(0)\;=\;u_0\ee
which received a rigorous interpretation within the framework of
stochastic analysis \cite{doob}. In the case when the
probability distribution of ${\bf u}(t)$, $t\to\infty$, is the
Maxwell one, 
$B(t)$ must be a Gaussian process, and the formula
(3) leads to the classical Ornstein-Uhlenbeck process.

Here,  we discuss properties of the process ${\bf u}(t)$,
and the corresponding process of displacements ${\bf x}(t)$, in
the case when $B\,=\,(B(t))_{t\geq 0}$ is
the Cauchy process, that is when $B$ satisfies the following
conditions:\\
\noindent
a) $B$ has independent increments, i.e. given $t_1<...<t_n$, the
differences $B(t_2)\,-\,B(t_1)$,
$B(t_3)\,-\,B(t_2)$,..., $B(t_n)\,-\,B(t_{n-1})$ are mutually independent
random variables,\\
b) $B$ has stationary increments, i.e. the probability distribution
of $B(t\:+\:\tau)\:-\:B(\tau)$ is
independent of $\tau$,\\
c) $B$ is continuous in probability, that is $\lim_{t\to s}B(t)\:=
\:B(s)$
in probability,\\
d) the characteristic function of $B$ is given by
$$E[e^{ipB(t)}]\;=\;e^{-t\psi(p)}$$
where $\psi(p)\,=\,\sigma^2|p|$.\\

All the above requirements form a mathematically consistent definition
of the Markovian jump-type  process in question, e.g. Cauchy process.
Notice that a suitable modification of the condition d) (set
$\psi(p)\,=\, - \alpha^2 p^2$ in the exponent; we refer to the
general form of the L\'{e}vy-Khintchine  formula) would leave us with
  the familiar Wiener process.

From a physical point of
view, solutions of induced partial (here, pseudo-) differential euqations
are  most important, and those incorporate transition probability densities
and densities of the process.

Notice that  the process of displacements is determined by ${\bf u}(t)$
in the standard way
\be {\bf x}(t)\;=\;{\bf x}(0)\;+\;
\int\limits_0^t{\bf u}(\tau)d\tau,\quad {\bf x}(0)\;=\;x_0\ee
Hence,  we should be able to derive  relevant densities  and transition
densities  not only for the  velocity process
but also for the   induced  spatial process.
 Additionally, if we wish to interpret
${\bf u}(t)$  as  a genuine velocity field for the process of spatial
displacements   ${\bf x}(t)$ (the mere formal attribution of the
velocity name to our random variable ${\bf u}(t)$ is highly misleading,
in view of an apparent discontinuity of sample paths), a careful
analysis of differentiability properties (in what sense ?) of
${\bf x}(t)$   is here necessary.

By integrating equation (2) we obtain that for $t\geq s$
\be {\bf u}(t)\;=\;\els{\bf u}(s)\;+\;\elt\cts e^{\lambda\tau}
dB(\tau)\ee
The integration of (3) yields
$${\bf x}(t)\;=\;{\bf x}(s)\;+\;\cts[e^{-\lambda(\tau\,-\,s)}
{\bf u}(s)\;+\;\ela\int\limits_s^{\tau}
e^{\lambda\beta}dB(\beta)]d\tau$$
$$=\;{\bf x}(s)\;+\;\llt{\bf u}(s)\;-\;
\cts d\tau(\frac{\ela}{\lambda})'\int\limits_s^{\tau}
e^{\lambda\beta}dB(\beta)$$

Integrating the last summand by parts and using the double
integration
formula involving a derivative with respect to the interior
integral, cf. (3.12) in \cite{doob}, we get
$$\cts d\tau(\frac{\ela}{\lambda})'
\int\limits_s^{\tau}e^{\lambda\beta}
dB(\beta)\;=\;\frac{\elt}{\lambda}
\cts e^{\lambda\tau}dB(\tau)\;-\;\frac{1}{\lambda}\cts dB(\tau)$$

Hence
\be {\bf x}(t)\;=\;{\bf x}(s)\;+\;\llt{\bf u}(s)\;+\;
\cts\frac{1\:-\:e^{-\lambda(t\,-\,\tau)}}{\lambda}
dB(\tau)\ee
which mimics (is identical with respect to the form)  a  fairly
traditional expression for a spatial random variable of the standard
Ornstein-Uhlenbeck process (with Wiener increments put instead of the
Cauchy  increments in the last summand).

\section{Probability densities and transition probability densities
for ${\bf u}(t)$   and ${\bf x}(t)$}

There is a number of (equivalent) procedures to deduce a probability
density of the process $ {\bf u}(t)$ from the Cauchy increments
statistics, see \cite{metzler,west}.
We shall follow a direct probabilistic route.

 In order to simplify the notation we write $P[X\,=\,x]$ for the
 density
 of the probability distribution of a random variable $X$, that is
 $P[X\in\Gamma]\,
=\,\int_{\Gamma}P[X\,=\,x]dx$ for $\Gamma\subset{\bf R}$.

 Suppose that  $f$ is a continuously differentiable function such
 that $f(\tau)\geq 0$,
and let $X\,=\,\cts f(\tau)dB(\tau)$. Terms of this functional
form  are encountered in formulas (4) - (6).

The random variable $X$ is the limit of the following sum
\be \sum\limits_{k=0}^{n-1}f(\tau_k)[B(\tau_{k+1})\:-\:B(\tau_k)]\ee
where $s\,=\,\tau_0<\tau_1<...<\tau_n\,=\,t$ is the partition of
the interval $[s,\,t]$. Because the process
$B$ has independent increments,  the probability density of (7) is
the convolution of densities of its
summands which, since the process has stationary increments, are
equal to
$$\frac{1}{\pi f(\tau_k)}\frac{\sigma^2(\tau_{k+1}\,-\,
\tau_k)}{(\frac{x}{f(\tau_k)})^2\:+\:\sigma^4
(\tau_{k+1}\,-\,\tau_k)^2}\;=\;\frac{1}{\pi}\frac{\sigma^2f(\tau_k)
\triangle\tau_k}{x^2\:+\:
(\sigma^2f(\tau_k)\triangle\tau_k)^2}$$

Because the Fourier transform maps the convolution to multiplication and
$$(\frac{1}{\pi}\frac{\sigma^2f(\tau_k)\triangle\tau_k}{x^2\:+\:
(\sigma^2f(\tau_k)\triangle\tau_k)^2})^{\wedge}(p)\;=
\;e^{-\sigma^2|p|f(\tau_k)\triangle\tau_k}$$
we get
 $$P[(\sum\limits_{k=0}^{n-1}f(\tau_k)[B(\tau_{k+1})\:-
 \:B(\tau_k)])\;=\;x]$$
$$=\;(\prod\limits_{k=0}^{n-1}e^{-
\sigma^2|p|f(\tau_k)\triangle\tau_k})^{\vee}(x)\;=\;
(\exp(-\sigma^2|p|\sum\limits_{k=0}^{n-1}f(\tau_k)
\triangle\tau_k))^{\vee}(x)$$
where $f^{\wedge}$ and $f^{\vee}$ denote the Fourier
transform and its inverse respectively.
By taking the
limit $n\to\infty$ we obtain that
$$P[X\,=\,x]\;=\;(\exp(-\sigma^2|p|\fta ))^{\vee}(x)\quad $$
and so the general formula
\be P[X\,=\,x]\;=\;\frac{1}{\pi}\frac{\sigma^2\fta}{x^2\:+
\:(\sigma^2\fta )^2}\ee
is valid. We shall exploit Eq. (8) repeatedly in below.   \\

{\bf Remark 1}: An apparent generalization of the previous
observation
is posssible. Assume that $B(t)$ is a L\'evy stable process with the
characteristic function
$$\psi(p,\,t)\;=\;\exp(-\sigma^2 t|p|^{\alpha}),\quad 0<\alpha\leq 2$$
Then, there holds  $$P[\cts f(\tau)dB(\tau)\;=
\;x]\;=\;(\exp[-\sigma^2(\cts f^{\alpha}(\tau)
d\tau)|p|^{\alpha}])^{\vee}(x)$$ \\

Presently we shall use the formula (8) to calculate transtion
probability densities of processes ${\bf u}(t)$  and
${\bf x}(t)$.

Let $f(\tau)\,=\,e^{-\lambda(t\,-\,\tau)}$. Then, by equation (5),
\be P[{\bf u}(t)\,=\,u|{\bf u}(s)\,=\,v]\;=\;\frac{1}{\pi}
\frac{\sigma^2(t\,-\,s)}{(u\,-\,v\els)^2\:+\:
\sigma^4(t\,-\,s)}\ee
where $\sigma^2(t\,-\,s)\,=\,\sil (1\:-\:\els)$, see e.g.
also \cite{west}.

Since ${\bf u}(0)\,=\,u_0$,
the probability density of ${\bf u}(t)$ is given by
\be P[{\bf u}(t)\,=\,u]\;=\;\frac{1}{\pi}
\frac{\sit}{(u\:-\:u_0\elt)^2\:+\:\sik}\ee

We now turn to the process ${\bf x}(t)$. Since ${\bf u}(s)$
is independent of $B(t)$ for all $t\geq s$
\cite{doob}, it is also independent of the integral
$\int_s^t f(\tau)dB(\tau)$.
Therefore, the probability distribution of the sum
$$\llt{\bf u}(s)\;+\;\cts\frac{1\:-\:e^{-\lambda(t\:-\:\tau)}}
{\lambda}dB(\tau)$$
is the convolution of its ingredients.

Let $f(\tau)\,=\,\frac{1\:-\:e^{-\lambda(t\:-\:\tau)}}{\lambda}$.
Because of
$$\fta\;=\;\frac{1}{\lambda^2}\lts$$
 by formula (8) there holds
\be P[\cts\frac{1\:-\:e^{-\lambda(t\:-\:\tau)}}{\lambda}dB(\tau)\:
=\:x]\;=\;\frac{1}{\pi}
\frac{(\sla)^2\lts}{x^2\:+\:(\sla)^4\lts^2}\ee

On the other hand, by (10), we  have
\be P[\llt{\bf u}(s)\:=\:u]\;=\;
\frac{1}{\pi}\frac{\sigma^2(s)a(t\,-\,s)}{(u\:-\:
u_0e^{-\lambda s}a(t\,-\,s))^2
\:+\:\sigma^4(s)a^2(t\,-\,s)}\ee
where $a(t\,-\,s)\,=\,\llt$.

The Fourier transform of (11) and (12) are equal to
\be \exp[-(\sla)^2\lts |p|]\ee
and
\be \exp[-\sigma^2(s)a(t\,-\,s)|p|]\exp[-iu_0
e^{-\lambda s}a(t\,-\,s)p]\ee
respectively.

Because of
$$(\sla)^2\lts\;+\;\sigma^2(s)a(t\,-\,s)\;=
\;(\sla)^2(\elt\:-\:e^{-\lambda s}\:+\:\lambda(t\:-\:s))$$
the mulitiplication of transforms (13), and (14)
followed by  taking the inverse Fourier transform  of the result,
gives us a transition probability density of the spatial
process
\be P[{\bf x}(t)\:=\:y|{\bf x}(s)\:=
\:x]\;=\;p(y,\,t|x,\,s)\;=\;\frac{1}{\pi}
\frac{g(t,\,s)}{(y\:-\:x\:-\:u_0h(t,\,s))^2\:+\:g^2(t,\,s)}\ee
where $$g(t,\,s)\;=\;(\sla)^2(\elt\:-\:e^{-\lambda s}\:+
\:\lambda(t\:-\:s))$$
and
$$h(t,\,s)\;=\;\frac{e^{-\lambda s}\:-\:\elt}{\lambda}$$

Finally, because ${\bf x}(0)\,=\,x_0$,  the probability
density of the process ${\bf x}(t)$ is given by
\be P[{\bf x}(t)\:=\:x]\;=\;\frac{1}{\pi}\frac{(\sla)^2
(\elt\:-\:1\:+\:\lambda t)}{(x\:-\:x_0\:-\:
u_0\frac{1\:-\:\elt}{\lambda})^2\:+\:
(\sla)^4(\elt\:-\:1\:+\:\lambda t)^2}\ee
Compare e.g. the corresponding formula for the displacements
of the standard (Wiener noise-supported) Ornstein-Uhlenbeck
process, \cite{chandra}.

\section{Properties  of  the process   ${\bf u}(t)$}

Considerations of the previous sections may leave us with
an impression that a construction  of the
Ornstein-Uhlenbeck process supported by  Cauchy noise
is in fact complete.  We have in hands not only  It\^{o}
type stochastic differential  equations that are amenable to
 direct computer simulations, \cite{weron,mantegna}, but
also explicit
expresions for probability densities and transition probability
 densities for both
processes:  ${\bf u}(t)$   and  ${\bf x}(t)$.   In case of Markov
processes   such data are known to specify the process uniquely,
\cite{lefever}.

However, some alarm bells need to switched on
at this point.   The standard (stationary)
Ornstein-Uhlenbeck velocity process is Markovian  (in the Gaussian
case the Ornstein-Uhlenbeck process is the \it only \rm
continuous in probability stationary Markov process, \cite{breiman}),
 but the induced (integrated) spatial process is  not Markovian.
Using  an explicit expression for
the transition probability density it is easy to verify that the
Chapman-Kolmogorov identity does not hold true.
Therefore,  Markov property
is normally attributed  to a two-component, phase-space
version of the Ornstein-Uhlenbeck process, \cite{chandra}.
In case of the Ornstein-Uhlenbeck-Cauchy process the situation
is somewhat different.

\subsection{Markovianess and  stochastic continuity}

First of all let us notice that
 ${\bf u}(t)$ is a time-homogeneous (but not stationary)
 Markov process.
 Markov property is clear from the  construction
 since Chapman-Kolmogorov identity  can be verified by
 inspection and it  is  a  classic observation that
 nonnegative and normalized  functions which
 obey the Chapman-Kolmogorov equation are
    necessarily Markovian transition   probabilities.

Since  the probability density (10) of the process depends explicitly
on time, our
process  ${\bf u}(t)$ is not stationary.

{\bf Remark 2}:
That needs to be
contrasted with the standard (Gaussian and stationary)
Ornstein-Uhlenbeck
 process features where the transition probability density
 is time-homogeneous, while the density of the process does
 not depend on time at all. Indeed (we consider one spatial
 dimension and utilize dimensional units) the transition density
 $$p(y,t|x,s) = (\gamma /2\pi D \{1 - exp[-2\gamma (t-s)]\})^{-1/2}
 \cdot exp(- {{\gamma \{x-yexp[-\gamma(t-s)]\}^2}\over
 {2D\{1-exp[-2\gamma(t-s)]\}}})$$
 with $s<t$, has an invariant  density  $\rho (x)=
 (\gamma /2\pi D)^{-1/2}\cdot exp(-\gamma x^2/2D)$. The drift of the
 process reads $b(x)=-\gamma x$ and $p$ solves the Fokker-Planck
 (second Kolmogorov)  equation
 $\partial _tp= D\triangle _xp - \nabla _x(bp)$. \\

Now, we shall demonstrate  an important property (mentioned before
in connection with the Ornstein-Uhlenbeck process)
 of the so-called stochastic continuity which is a necessary
 condition
 to give a stochastic process an unambigous status,
 \cite{lefever,gihman,olk0}.
 Namely, we need to show that
 for any $\epsilon>0$ the following equation is satisfied
\be \lim\limits_{t\to s}P[|{\bf u}(t)\:-\:{\bf u}(s)|\:
\geq\:\epsilon]\;=\;0\ee

This equation is equivalent to
\be \lim\limits_{t\to 0}\int\limits_{|u-v|
\geq\epsilon}p_t(u|v)du\;=\;0\ee
Because of
$$\int\limits_{|u-v|\geq\epsilon}p_t(u|v)du\:=\:1\:-\:\frac{1}{\pi}
[\arctan\frac{\epsilon\:+\:v(1\:-\:\elt)}{\sit}\:+
\:\arctan\frac{\epsilon\:-\:v(1\:-\:\elt)}{\sit}]$$
and remembering that $\sigma^2(t)\,=\,\sil (1\:-\:\elt)$,
the stochastic continuity property does follow.

It is perhaps not useless to emphasize  that in typical  Gaussian
process investigations, stochastic continuity of the process is
a necessary (but still insufficient) condition for the process to have
continuous sample paths. Hence it is always explictly mentioned
in the context of diffusion pocesses, \cite{lefever}.
The Cauchy noise-supported process is surely not diffusive and its
trajectories are discontinuous (jump-type) paths,
\cite{klaud,olk}).

\subsection{Local moments in the Cauchy case: forward drift --
the existence issue}

The nonexistence   of moments of the probability measure in case of
the Cauchy process is another source of difficulties, since the
standard local characteristics of the diffusion-type process like the
 drift and the diffusion function (or coefficient)   seem to be excluded
 in the present case.

However, for  the considered Ornstein-Uhlenbeck-Cauchy process,
the notion of the forward drift of the process proves to make
sense (!).
We shall first  discuss the drift issue for the  process
${\bf u}(t)$.

Let us start with the following definition. Suppose
$p(y,\,t|x,\,s)$, $t\geq s$, is a Markov transition function and
let $X_t$ be the associated Markov
process.
Guided by the analogy with diffusion processes we say that
the process $X_t$ has a drift (in fact, forward drift) if the
following limit
\be \lim\limits_{t\to s}\frac{1}{t-s}
\int\limits_{|y-x|\leq\epsilon}(y\:-\:x)p(y,\,t|x,\,s)dy\ee
does not depend on the choice of $\epsilon>0$.
If so, then its value depending only on $(x,\,s)$ we
denote by $b(x,\,s)$ and call it the drift coefficient.

Clearly, if $p$ is homogeneous in time, then
the drift coefficient depends only on the  variable $x$.
Let us emphasize that in the above definition we do not
require the process $X_t$
 to have finite moments.\\

We claim that the jump-type Markov  process ${\bf u}(t)$ has
a (forward) drift which reads  $b(v)\,=\,-\lambda v$.\\

 At first we calculate the indefinite integral
$$I\;=\;\frac{1}{\pi}\int (u\:-\:v)
\frac{\sit du}{(\yks)^2\:+\:\sik}$$

Substituting $z\,=\,\yks$, we rewrite that integral  as
$$\frac{\sit}{\pi}\int\frac{zdz}{\zes}\;+\;
\frac{v}{\pi}(\elt\:-\:1)\int\frac{\sit dz}{\zes}$$
$$=\;\frac{\sit}{2\pi}\log (\zes)\;+\;
\frac{v}{\pi}(\elt\:-\:1)\arctan (\frac{z}{\sit})$$
Hence
$$I\;=\;\frac{\sit}{2\pi}\log[(\yks)^2\:+\:\sik]\;+
\;\frac{v}{\pi}(\elt\:-\:1)\arctan[\frac{\yks}{\sit}]$$
and consequenly the limit
$$\lit I|^{u=v+\epsilon}_{u=v-\epsilon}\;=$$
$$\lit\frac{\sit}{2\pi}(\log[(v\:+
\:\epsilon\:-\:v\elt)^2\:+\:\sik]\;-\;
\log[(v\:-\:\epsilon\:-\:v\elt)^2\:+\:\sik])$$
$$+\;\lit\frac{v}{\pi}(\elt\:-\:1)(\arctan[\frac{v\:+
\:\epsilon\:-\:v\elt}{\sit}]\;-\;
\arctan[\frac{v\:-\:\epsilon\:-\:v\elt}{\sit}])$$
$$=\;0\;-\;\lambda\frac{v}{\pi}(\frac{\pi}{2}\:+
\:\frac{\pi}{2})\;=\;-\lambda v\quad $$
exists and is $\epsilon $-independent.
This is the forward drift of
the process ${\bf u}(t)$ which  proves a consistency of
the derived
transition probability  density with the stochastic
differential equation (5).

To our knowledge, such consistency check has
never been performed before in discussions of L\'{e}vy flights
and anomalous diffusion processes.

On the other hand, it is well
known that for Markovian diffusion pocesses all local
characteristics of motion (conditional expectation values
that yield drifts and variances)   are derivable from
transition probability densities, supplemented (if needed)
 by the density of the process, cf. \cite{olk0}.
We have demonstrated that, in the non-Gaussian  context,
 the nonexistence of moments
does not necessarily imply   the nonexistence of local
characteristics (drifts) of the process.

As a consequence, once a formal definition is adopted of
a stochastic differential equation whose deterministic
driving term (functionally unrestricted drift) is subject
to perturbations by L\'{e}vy   flights, the process may still
posess
local characteristics (forward drift) that are  in turn
derivable by
means of its transition density.  Our derivation in the Cauchy
noise case   is limited to linear functions
of random variables (linear systems, \cite{west,metzler}.
Possible generalizations to stochastic differential equations with
driving terms represented by
nonlinear and possibly time-dependent functions need to be carefully
examined.

This is an uncomfortable situation, since
a  formal computer experimentation  may not
indicate any inconsistency of  the formalism.
Even worse, the uncommented visualization
 effectively may convey  misleading  or entirely
 wrong messages if uncritically accepted.
(The  rigorous existence theorems
available in the mathematical literature pertain to linear
systems as well, \cite{chojnowska,taqqu} and extend to
perturbations by general L\'{e}vy processes.)

\subsection{Markov generators and  Kolmogorov
 (Fokker-Planck type) equations}

Once densities and transition probability densities  have been
obtained from the first principles, we can invert the problem
(that is a commonly shared
 viewpoint in the physics-oriented  research) and ask for
 differential (evolution) equations obeyed by them.
The Fokker-Planck equation is an obvious example in case
of Markovian diffusion processes, while various forms of the
Master equation  were adopted to extend the standard
jump-processes (Poisson or more generally-point processes)
framework to more singular  step or jump-type    ones.

In the case of unperturbed (free)
L\'{e}vy processes basically all interesting (covering stable laws of
probability) evolution equations were classified by means of Fourier
transform techniques, \cite{west,monti,metzler,uchaikin,olk}.
A disregarded point was that in case of Markov processes a single
(Fokker-Planck or Master equation - type) evolution equation does
not characterise the  process uniquely.  Both forward and backward
evolution equations need here to be involved, cf.
\cite{breiman,olk0}.  Except for Refs. \cite{west,metzler} no
attempt was made  to investigate such equations for
deterministically driven L\'{e}vy systems.

To elucidate that issue, we shall
next consider the generator of a Markov transition function
$p_t(y|x)$ for the Cauchy-perturbed process (cf. the previous Section).

Let us recall that it is  defined by
\be (Lf)(x)\;=\;\lit [\caa p_t(y|x)f(y)dy\;-\;f(x)]\ee
where the domain of definition consists of all functions
$f\in C_0({\bf R})$, whose limit on the right hand 
side in (19) exists uniformly with respect to the variable $x$.

It is worth noting that when the transition
function is stochastically continuous (see the previous section),
then the corresponding semigroup $T_t$ in
$C_0({\bf R})$ defined by
\be (T_tf)(x)\;=\;\caa p_t(y|x)f(y)dy\ee
is strongly continuous, and so its generator $L$ is
densely defined.

In such a case we can also define an
adjoint semigroup $T_t^*$ acting on the space of
(probability) densities $L^1({\bf R},\,dx)$,
\be (T_t^*\rho)(u)\;=\;\caa p_t(u|v)\rho(v)dv\ee
Its generator we denote by $L^*$.\\

Arguments of the present section involve a little bit of
a  mathematical formalism to stay in conformity with the
classic work of Feller and Dynkin on evolution equaitons for
Markov processes, cf. \cite{olk,olk0} for references.

 Suppose $L$ is
the generator of the semigroup associated with the process
${\bf u}(t)$ and let $L^*$ be its adjoint.

We wish to demonstrate that
\be L\;=\;L_0\;+\;b\nabla \ee
and
\be L^*\;=\;L_0\;-\;\nabla (b\,\cdot ) \ee
where $L_0$ is the generator of the
Cauchy process $B$ (we have used an explicit
 notation  $L_0=|\nabla | $  in Refs. \cite{klaud,olk},
 see also \cite{metzler}) and $b(v)\,=\,-\lambda v$.\\

To this end, we first   observe that for $p_t(u|v)$ given by
the formula (9) the associated semigroup $T_t$ maps $C_0({\bf R})$
to $C_0({\bf R})$.
Since $p_t(u|v)$ is stochastically continuous, $T_t$ is
strongly continuous.

Next,
we calculate the Fourier transform of equation (20).
$$(Lf)^{\wedge}(p)\;=\;\lit [\caa\caa\frac{1}{\pi}
\frac{\sit}{(\yks)^2\:+\:\sik}f(u)e^{-ipv}dudv\;-\;\fep]$$

Substituting $z\,=\,v\elt\:-\:u$, $dv\,=\,e^{\lambda t}dz$,
we obtain that
$$(Lf)^{\wedge}(p)\;=$$
$$\lit [e^{\lambda t}\caa dz\frac{1}{\pi}\exp (-ize^{\lambda t}p)
\frac{\sit}{\zes}\caa du \exp (-iue^{\lambda t}p)f(u)\;-\;\fep]$$
$$=\;\lit [e^{\lambda t}\exp (-\sit e^{\lambda t}|p|)
\hat{f}(e^{\lambda t}p)\;-\;\fep]$$
$$=\;\lit [e^{\lambda t}\exp (-\sit e^{\lambda t}|p|)\:-\:1]
\hat{f}(e^{\lambda t}p)\;+\;\lit
[\hat{f}(e^{\lambda t}p)\:-\:\fep]$$
$$=\;-\sigma^2|p|\fep\;+\;\lambda\fep\;+\;\lambda p\hat{f}'(p)$$

By taking the inverse Fourier transform and using the identity
$$(p\hat{f}')^{\vee}(v)\;=\;\frac{1}{2\pi}
\caa e^{ipv}p\hat{f}'(p) dp\;=\;-\;\frac{1}{2\pi}\caa (e^{ipv}p)'
\fep dp$$
$$=\;-f(v)\;-\;iv(p\hat{f})^{\vee}(v)\;=\;-f(v)\;-\;vf'(v)$$
we arrive at
$$Lf(v)\;=\;L_0f(v)\;-\;\lambda vf'(v)$$
where $L_0\,=\,-\sigma^2|\nabla|$.

Hence $L\,=\,L_0\,+\,b\nabla$.

Because of
$$L^*(\rho)(u)\;=\;\lit [\caa p_t(u|v)\rho(v)dv\;-\;\rho(u)]$$
so, by similar calculations as above, we obtain
$$L^*(\rho)(u)\;=\;L_0\rho(u)\;-\;\nabla(b(u)\rho(u))$$
That ends the demonstration. \\

As a consequence, since $T_t:\,C_0({\bf R})\to C_0({\bf R})$ is
strongly continuous, almost all paths of the
process ${\bf u}(t)$ are {\it cadlag}, that is they are continuous
from the right and have finite left-hand limits (see chap.II
sec.4 in \cite{gihman},  vol.II).
There follows also from  Eqs. (23 ) and (24)  that the  transition
probability function of the process ${\bf u}(t)$ satisfies
the backward equation 
\be \frac{\partial p_t(u|v)}{\partial t}\;=\;
L_0p_t(u|\cdot)(v)\;+\;b(v)\nabla_vp_t(u|v)\ee
and the forward equation (the Fokker-Planck equation analogue)
\be \frac{\partial p_t(u|v)}{\partial t}\;=\;
L_0p_t(\cdot|v)(u)\;-\;\nabla_u[b(u)p_t(u|v)]\ee
for  definitions see, for example, chap. 15 sec. 4 in
\cite{breiman}.

It is trivial to check by inspection that the transition probability
function of the (free) Cauchy process obeys both those equations
with   $b(u)$   set equal identically $0$.

\subsection{Asymptotics: ergodicity, mixing, exactness}

We have in hands an explicit expression for the density of the
process ${\bf u}(t)$. One of efficient ways to investigate the
complexity of the  involved random  dynamics is not necessarily
via a direct recourse to  sample paths, but rather via
studying  asymptotic properties  of probability densities,
cf. \cite{lasota}.

Let us  consider the asymptotic properties of the process
${\bf u}(t)$.
By direct calculations we check that the density
$$\rho_0(u)\;=\;\frac{1}{\pi}\frac{\sini}{u^2\:+\:\sici}\;=\;
\frac{1}{\pi}\frac{\sigma^2/\lambda}{u^2\:
+\:\sigma^4/\lambda^2}$$
is stationary with respect to the dual semigroup $T_t^*$.
Therefore, irrespectively of its initial probability
distribution, $P[{\bf u}(\infty)\:=\:u]\:=\:\rho_0(u)$.   \\

{\bf Remark 3}:
Normally, if we get a convergence of a density  in the asymptotic
($t \rightarrow \infty $) regime  to a unique density, we say
about  an asymptotic stability. In case when for  every initial
density  we get a set spanned by a finite  number of densities,
we say about an asymptotic periodicity.
We may  also have  a situation that every initial density is
dispersed under the action of a Markov operator. That is
related to the concept of  sweeping.\\

Our  knowledge of the explicit formula of the
transition probability density for the semigroup $T_t^*$
allows us to examine its ergodic properties in a more
detailed way.   For example,  a point-wise convergence of the
Cauchy process transition density   to the stationary Cauchy
probability density was ivestigated in \cite{west}.

Let us recall (see \cite{lasota} for the
definition and more details) that a Markov semigroup
$T_t^*$ is mixing, if for any density $\rho$,
$T_t^*\rho$
tends to a stationary density in a weak sense,
and exact, if this limit holds in the $L^1$-norm.

Hence exactness is a stronger property and implies mixing,
ergodicity  of the dynamics being a straightforward consequence.

(It mighth be  worth noting that strong  mixing properties of
 the standard Ornstein-Uhlenbeck velocity fields were
 discussed and visualized  by computer simulations in
 \cite{carmona}. The point-wise convergence of the probability
 density of the process to its stationary limit was established
 in \cite{namiki}.)

The dynamics  induced by Cauchy noise (and other stable noises)
shows higher level of complications and is not only mixing, but
also exact.  We shall provide a demonstration of this property
in the spirit  of Ref. \cite{lasota}.

In fact, what we claim is that  $T_t^*$ is exact (hence
both mixing and ergodic).

 Since $\rho_0$ is a stationary density we have to show that
\be \lini\|\rho_t\:-\:\rho_0\|_1\;=\;0 \ee
where $\rho_t(u)\,=\,\int p_t(u|v)\rho(v)dv$  and $\rho (v)$
is an arbitrary initial density.
To this end we need an auxiliary lemma which comprises the most
technical segment of the paper.\\

{\bf Lemma }   \\
For $t\to\infty$, $\|p_t(\cdot |v)\:-\:\rho_0\|_1\to 0$ uniformly in
$v$ on compact sets.\\

{\bf Proof}: We shall show that $\forall N\in{\bf N}\;\forall
\epsilon>0\;\exists t_0>0$ such that
$$\forall t>t_0\;\forall v\in [-N,\,N]\;\;
\|p_t(\cdot |v)\:-\:\rho_0\|_1\:<\:\epsilon$$ 

Let us begin from
$$\|p_t(\cdot |v)\:-\:\rho_0\|_1\:=\:\frac{1}{\pi}\caa
|\frac{\sit}{(\yks)^2\:+\:\sik}\;-\;
\frac{\sini}{u^2\:+\:\sici}|du$$
$$\leq\;\frac{1}{\pi}\caa |\frac{|\sit\:-\:
\sini|}{(\yks)^2\:+\:\sik}du\;+$$
$$\sip\caa |\frac{1}{(\yks)^2\:+\:\sik}\;-\;
\frac{1}{u^2\:+\:\sik}|du$$
$$+\;\sip\caa |\frac{1}{u^2\:+\:\sik}\;-\;
\frac{1}{u^2\:+\:\sici}|du\;=\;\frac{\sini\:-\:\sit}{\sit}$$
$$+\;\sip\caa \frac{|u^2\:-\:(\yks)^2|}{[(\yks)^2\:+\:
\sik][u^2\:+\:\sik]}du\;+$$
$$\frac{\sini[\sici\:-\:\sik]}{\pi}\caa\frac{du}{[u^2\:+\:
\sik][u^2\:+\:\sici]}$$

The first summand, denoted by $I_1$, equals to $\elt/(1\:-\:\elt)$
and so is less than $\epsilon/3$ for all
$t>t_1$,  provided $t_1=\,\frac{1}{\lambda}
\log(\frac{3}{\epsilon}\:+\:1)$.

For $t\geq\frac{\log 2}{\lambda}$ we have
$\sik\geq\frac{1}{4}\sici$ and so the third summand, denoted by
$I_3$, can be estimated as follows
$$I_3\;\leq\;\sip [\sini\:+\:\sit][\sini\:-\:\sit]
\caa\frac{du}{[u^2\:+\:(\frac{\sini}{2})^2]^2}$$
$$\leq\;\frac{2\sigma^6(\infty)\elt}{\pi}\cdot
\frac{4\pi}{\sigma^6(\infty)}\;=\;8\elt$$

Therefore, for any $t>t_3=\:\mbox{max}(\frac{1}{\lambda}\log 2,\,
\frac{1}{\lambda}\log\frac{24}{\epsilon})$ ther holds
$I_3\leq\epsilon/3$.

Finally, we estimate the second summand, denoted by  $I_2$.
At first let us notice that
$$I_2\;=\;\sip\caa\frac{|v|\elt|2u\:-\:v\elt|}{[(\yks)^2\:+\:
\sik][u^2\:+\:\sik]}du$$
$$=\;8|v|\elt\sip\caa\frac{|x|dx}{[(x\:-\:v\elt)^2\:+\:
4\sik][(x\:+\:v\elt)^2\:+\:4\sik]}$$
where $x\,=\,2u\,-\,v\elt$.

Hence, for all $t\geq t_4=\,\frac{1}{\lambda}\log(1\:+\:
\frac{N}{2\sini})$
there holds  $|v|^2e^{-2\lambda t}\leq 4\sik$ for all
$v\in [-N,\,N]$, and so
$$I_2\;\leq\;8\elt\sip\caa\frac{|x|dx}{x^4\:+
\:(4\sik)^2}\;=\;\frac{1}{\sini}\frac{\elt}{(1\:-\:\elt)^2}$$

Therefore, $I_2<\,\epsilon/3$ for all $t>t_2=\:
\mbox{max}(t_4,\,t_5)$, where $t_5$ is determined by
$3e^{-\lambda t_5}\,=\,\sini(1\:-\:e^{-\lambda t_5})^2\epsilon$.
Thus $\|p_t(\cdot |v)\:-\:\rho_0\|_1<\epsilon$
for all $t>t_0=\:\mbox{max}(t_1,\,t_2,\,t_3)$ what ends proof
of the lemma.$\Box$ \\

Now, we are ready to adddress the exactness issue for  $T_t^*$.

We observe that
$$\|\rho_t\:-\:\rho_0\|_1\;=\;\caa |\caa p_t(u|v)\rho(v)dv\:-
\:\rho_0(u)|du\;=$$
$$\caa |\caa(p_t(u|v)\:-\:\rho_0(u))\rho(v)dv|du\:
\leq\:\caa\|p_t(\cdot |v)\:-\:\rho_0\|_1\rho(v)dv$$
$$\leq\:\int\limits_{-N}^N\|p_t(\cdot |v)\:-\:\rho_0\|_1
\rho(v)dv\;+\;2\int\limits_{[-N,\,N]^c}\rho(v)dv$$
where $[-N,\,N]^c$ denotes the complement of the interval $[-N,\,N]$.

Let us  choose $N$ such that the
second integral in the above is less than $\epsilon/4$, and next,
in conjunction with  Lemma, we choose  $t_0$ such that for all
$t>t_0$,  there holds
$\|p_t(\cdot |v)\:-\:\rho_0\|_1<\epsilon/2$ for all $v\in [-N,\,N]$.

Then
$$\int\limits_{-N}^N\|p_t(\cdot |v)\:-\:
\rho_0\|_1\rho(v)dv\:\leq\:\frac{\epsilon}{2}$$
and so $\|\rho_t\:-\:\rho_0\|_1<\,\epsilon$, which complets the
exactness demonstration for $T_t^*$. \\

{\bf Remark 4}:
In our  considerations of the exactness issue  the
 stationary Ornstein-Uhlenbeck-Cauchy process has been employed.
 This
process is a direct   L\'{e}vy  stable  analogue
 of the standard Gaussian Ornstein-Uhlenbeck process .
 There is however an important difference,  \cite{taqqu}.
  In the Gaussian case all stationary Markov processes
are Ornstein-Uhlenbeck (which in turn is unique).
In particular,  the standard  process
coincides  with  its reverse (fully anticipating) version.
It turns out that in the Cauchy
case there exist at least two different stationary Markov processes,
since  the reverse one does not coincide with the forward
(nonanticipating)  one.
This  means that the so-called statistical inversion of the Markovian
 dynamics (cf. a discussion and references in the closing section
 of Ref. \cite{gar}), in case of stable L\'{e}vy processes,  makes
 a distiction between the "time arrow" direction. See e. g.
also time reversal and time adjointness problems encountered
in Refs. \cite{klaud,olk}.
Presumably all that  derives from the exactness property.
In modern attempts to devise a theoretical framework for
 nonequilibrium  statistical physics, based on exploitation
 of clasically chaotic systems,  most
 irregular and "most irreversible" dynamical phenomena are known
 to be generated by exact systems. \\

  We could as well  consider Eq. (1) with a
nondissipative $\lambda<0$ factor.
However, then the asymptotic properties of
the semigroup $T_t^*$ change in an essential way. Indeed, since
now the transition probability density
can be written as
$$p_t(u|v)\;=\;\frac{1}{\pi}\frac{\frac{\sigma^2}
{|\lambda|}(e^{|\lambda|t}\:-\:1)}
{(u\:-\:ve^{|\lambda|t})^2\:+\:(\frac{\sigma^2}
{|\lambda|}(e^{|\lambda|t}\:-\:1))^2}$$
then for any $N\in{\bf N}$ and any density $\rho$ there holds
$$\lini\int\limits_{-N}^Np_t(u|v)\rho(v)dv\;=\;0$$
In consequence  $T_t^*$ is sweeping, \cite{lasota}.
It means that $T_t^*$ has no stationary density and,
in consequence, there is no probability law at all for
the $t\to\infty$ limit of the process ${\bf u}(t)$.       \\

\section{ Properties of the process ${\bf x}(t)$}

\subsection{Markovianess and forward drift}

In case of the classic Ornstein-Uhlenbeck process, it is well
known that the spatial random variable does not represent
a Markov process.

A little bit surprisingly, in the  present (Cauchy noise) case, we can
prove that ${\bf x}(t)$ is a Markov process which is (like the
previous ${\bf u}(t)$) stochastically continuous.
  Moreover, while being a discontinuous process, nonetheless
  it has a forward drift equal  $b(s)\,=\,u_0e^{-\lambda s}$.\\

 To show  Markov property it suffices to check the Chapman-Kolmogorov
 equation for the transition function given by equation (15).
That immediately follows due to the additivity properties of
functions $g(t,\,s)$ and $h(t,\,s)$ which enter the formula for $p(y,\,t|x,\,s)$
$$g(t,\,t')\;+\;g(t',\,s)\;=\;g(t,\,s)$$
$$h(t,\,t')\;+\;h(t',\,s)\;=\;h(t,\,s)$$

The stochastic continuity can be shown by a direct verification of the
formula (18) which, in the nonstationary
case, reads
$$\lim\limits_{t\to s}\int\limits_{|y-x|\geq\epsilon}
p(y,\,t|x,\,s)dy\;=\;0$$

Also,  by direct calculations we check that the limit
$$\lim\limits_{t\to s}\frac{1}{t\:-\:s}
[\int\limits_{|y-x|\geq\epsilon}(y\:-\:x)p(y,\,t|x,\,s)dy]$$
does not depend on the chosen  $\epsilon$ cutoff, and
equals $u_0e^{-\lambda s}$. \\

{\bf Remark 5}: It is worth pointing out that  Markov property
of the pure spatial process ${\bf x}(t)$ is a distinguishing
feature of the Cauchy noise.
It does not hold for other $\alpha $-stable L\'evy processes,
in particular for
 the Gaussian one (the standard Ornstein-Uhlenbeck process).
The reason of this exception is rooted  in a particularly simple
form of the probability distribution of the process
$\int_s^tf(\tau)dB(\tau)$ when $\alpha=1$, see e.g. our Remark 1. \\

In contrast to the velocity process  our spatial process is no
longer time-homogeneous.
In the inhomogeneous  case instead of a one-parameter semigroup we
have a two-parameter family of operators
$T_{t,s}$ defined by
\be (T_{t,s}f)(x)\;=\;\caa p(y,\,t|x,\,s)f(y)dy\ee
which satisfy the composition rule $T_{t,t'}T_{t',s}\,=\,T_{t,s}$.

Therefore, we can introduce a time
dependent generator by the following formula
\be (M(s)f)(x)\;=\;\lim\limits_{t\to s}\frac{1}{t\:-\:s}
[\caa p(y,\,t|x,\,s)f(y)dy\;-\;f(x)]\ee

In analogy with our previous considerations, we can readily
identify an explicit  form of the generator M(s). Namely,
let us  assume that  $p(y,\,t|x,\,s)$ is the transition
function of the process ${\bf x}(t)$ and
$T_{t,s}$ are operators associated with this function.
Then
\be M(s)\;=\;-\sigma^2(s)|\nabla |\;+\;b(s)\nabla\ee
where $\sigma^2(s)$ is as (8) and $b(s)\,=\,u_0e^{-\lambda s}$.\\

Because the major steps of the demonstration  are essentially
the same as in the case of the process ${\bf u}(t)$, we skip them here.
\\

Furthermore, let us notice that in view of
 $\lim_{s\to\infty}M(s)\,=\,\frac{\sigma^2}{\lambda}|\nabla |$
 so, for large $t$ (i. e. asymptotically)
the process ${\bf x}(t)$ converges to the Cauchy process with
the transition function given by
$$p_t(y|x)\;=\;\frac{1}{\pi}\frac{t\sigma^2/\lambda}
{(y\:-\:x)^2\:+\:t^2\sigma^4/\lambda^2}$$

Hence, the dissipation constant $\lambda >0$ does the job "as usual",
though with no recourse to the standard Maxwell-Boltzmann notion of
thermal equilibrium and fluctuation-dissipation theorems.\\

\subsection{Sample paths features}

Let us  turn to a brief  discussion of the properties of sample
paths of the process ${\bf x}(t)$.
(We remember that sample paths of the jump-type process ${\bf u}(t)$
were {\it cadlag}.)

In his seminal
paper, \cite{doob},  Doob proved  that the displacements of
the standard Ornstein-Uhlenbeck
process satisfy
\be \lim\limits_{t\to\infty}\frac{{\bf x}_{OU}(t)\:-\:
{\bf x}_{OU}(0)}{t}\;=\;0\ee
almost surely, i.e. the above limit holds for almost all sample paths
(with probability 1).
That is interpreted as an ergodic theorem  applied to the velocity
process to give the strong law of large numbers, \cite{doob}, and
at the same time  as a statement about
the (sample) path of a single particle.

In the final remark on p.369, he also concluded that Eq. (31)
holds true also in the case when the noise $B$ is a stable
process with the characteristic
$\mu\geq 1$.
Hence, for the Cauchy process as well.

However, this conjecture appears  to be wrong, in view of the
 estimates we present in below.

 Because of
$$P[|{\bf x}(t)\:-\:{\bf x}(0)|\:>\:\epsilon t]\;=
\;\int\limits_{|x|>\epsilon}\frac{1}{\pi}
\frac{g(t,\,0)dx}{(x\:-\:u_0h(t,\,0))^2\:+\:g^2(t,\,0)}$$
$$=\;1\;-\;\frac{1}{\pi}[\arctan\frac{\epsilon t\:+
\:u_0h(t,\,0)}{g(t,\,0)}\:+\:
\arctan\frac{\epsilon t\:-\:u_0h(t,\,0)}{g(t,\,0)}]$$
we have
 $$\lim\limits_{t\to\infty}P[\frac{|{\bf x}(t)\:-
\:{\bf x}(0)|}{t}\:>\:\epsilon]\;=\;1\:-\:
\frac{2}{\pi}\arctan\frac{\epsilon\lambda}{\sigma^2}\:>\:0$$
Therefore, $\frac{{\bf x}(t)\:-\:{\bf x}(0)}{t}$ does not tend
to zero even in probability.

That  means that
sample paths of the process ${\bf x}(t)$ diverge to infinity
faster than time $t$ (up to dimensional constants).\\

{\bf  Remark 6}:
(i) We can generalize slightly the discussion and allow
the parameter $\lambda$ to depend on time
$$d{\bf u}(t)\;=\;-\lambda(t){\bf u}(t)dt\;+\;dB(t)$$
where $\lambda(t)$ is a continuous function. Then,
by integrating the above equation, we obtain
$${\bf u}(t)\;=\;b(t,\,s){\bf u}(s)\;+\;\cts b(t,\,\tau)dB(\tau)$$
where $b(t,\,s)\,=\,\exp[-\cts\lambda(\tau)d\tau]$.
By invoking our previous arguments it is easy to find
the probability distribution and
transition function for the  new process ${\bf u}(t)$
(and the new process of displacements).\\
We can also consider an $n$-dimensional situation, when $\vec{B}(t)$
is an ${\bf R}^n$-valued Cauchy process,
and the Langevin equation (1) is replaced by the following one
\be d\vec{u}(t)\;=\;-A\vec{u}(t)dt\;+\;d\vec{B}(t) \ee
where $A$ denotes now $n\times n$ matrix with real coefficients.
The existence of the solution for (32) in
a general setting of $\cal H$-valued processes, $\cal H$ being
a real and separable Hilbert space, was
established in
\cite{chojnowska}.\\
(ii) Sample paths of the process ${\bf x}(t)$ are also {\it cadlag}.\\

\subsection{Interpreting ${\bf u}(t)$ as a velocity variable for
${\bf x}(t)$: Limitations}

Finally, we shall discuss the relation between the process
of velocities ${\bf u}(t)$ and the process of
displacements ${\bf x}(t)$.
In the Ornstein-Uhlenbeck case the process $\uout$ is continuous
in the mean square, that is
$$\lim\limits_{h\to 0}E[|\uou(t\,+\,h)\:-\:\uou(t)|^2]\;=\;0$$
That  follows from the continuity of its covariance function
$$(t_1,\,t_2)\to E[\uou(t_1)\uou(t_2)]\;=\;
\sigma^2e^{-\lambda|t_1\,-\,t_2|}$$

Therefore, ${\bf x}_{OU}(t)$ exists as a limit in mean square
of the corresponding Riemann sums. Moreover,
since sample paths of $\uout$ are continuous, the integral
exists also almost surely and they both
coincide.
Hence ${\bf x}_{OU}(t)$ is not only mean square differentiable
but is also differentiable
in the sense of conventional mathematical analysis and its
derivative is $\uout$.

For the Cauchy process $B(t)$ the situation is different.
Because the moments of ${\bf x}(t)$ do
not exist so ${\bf x}(t)$ is not even continuous in mean square.
Moreover, since its sample paths are not
continuous, they have no derivatives   either.

However, ${\bf x}(t)$ \it has \rm a velocity field ${\bf u}(t)$
in a probabilistic sense.

Indeed, because ${\bf u}(t)$ is stochastically continuous
so for $h\to 0$   there holds
$$\frac{{\bf x}(t\:+\:h)\:-\:{\bf x}(t)}{h}\:=
\:\frac{1}{h}\int\limits_t^{t+h}{\bf u}(\tau)d\tau\:\to\:
{\bf u}(t)$$ in probability, compare e.g. Eq. (17).

That demonstrates that our naming (a priori) of ${\bf u}(t)$
the velocity random variable, can still be maintained to
a certain  extent, once
we pass to the induced spatial variable and try to recover back a
built-in information about the velocity process.
This feature is slightly  amusing and somewhat counterintuitive, since
both the Cauchy noise-supported velocity process and the induced
process of spatial displacements  are discontinuos with probability 1.
Anyway, the notion of velocity in the standard Ornstein-Uhlenbeck
process has its own limitations as well: its nondifferentiability
and thus the nonexistence of accelerations is not resolved but
merely bypassed by invoking the white noise calculus.

One should  perhaps recall at this point
 that our theoretical framework reduces to  a stochastic modelling
 of physical phenomena.
 That  constitutes a  fine-tuned  approximation,  in terms of
 stochastic processes, of
a generically  robust Reality.
Surely, we have not attempted her   genuine reproduction.
The obtained description is   much too detailed and thus
necessarily involves a number of  artefacts (the
nonexistenece of  accelerations in the standard Ornstein-Uhlenbeck
process is  one of them, an unbounded variation of the Wiener process
sample paths just another).
\\

{\bf Acknowledgement}:  Both authors receive support from the KBN
research grant No 2 P03B 086 16.

\end{document}